\documentclass{iopart}
\usepackage{amssymb,amstext}
\usepackage{graphicx,color}
\usepackage{cite}
\usepackage{bm}
\usepackage{soul}

\newcommand{\eqref}[1]{(\ref{#1})}
\newcommand{\vect}[1]{\bm{#1}}
\newcommand{\mean}[1]{\langle #1\rangle}

\newcommand{\id}{\mathbf 1}
\newcommand{\dm}{d^{\text{max}}}

\newcommand{\ra}{\rightarrow}
\newcommand{\al}{\alpha}
\newcommand{\gam}{\gamma}
\newcommand{\vhi}{\varphi}
\newcommand{\Qr}{Q_{l\backslash\gam}}
\newcommand{\Cn}{\mathcal C}
\renewcommand{\ij}{(i\ra j)}

\begin{document}

\title[Cycle representatives for the coarse-graining of driven systems]{Cycle representatives for the coarse-graining of systems driven into a non-equilibrium steady state}

\author{Fabian Knoch and Thomas Speck}
\address{Institut f\"ur Physik, Johannes Gutenberg-Universit\"at Mainz,
  Staudingerweg 7-9, 55128 Mainz, Germany}

\begin{abstract}
  A major current challenge poses the systematic construction of coarse-grained models that are dynamically consistent, and, moreover, might be used for systems driven out of thermal equilibrium. Here we present a novel prescription that extends the Markov state modelling approach to driven systems. The first step is to construct a complex network of microstates from detailed atomistic simulations with transition rates that break detailed balance. The coarse-graining is then carried out in the cycle space of this network. To this end we introduce the concept of representatives, which stand for many cycles with similar properties. We show how to find these cycle communities using well-developed standard algorithms. Removing all cycles except for the representatives defines the coarse-grained model, which is mapped back onto a network with far fewer states and renormalized transition rates that, however, preserve the entropy production of the original network. Our approach is illustrated and validated for a single driven particle.
\end{abstract}



\section{Introduction}

The reduction of degrees of freedom is arguably the most crucial step in computational sciences. Numerical models can be formulated at several levels of detail: from \emph{ab initio} including the electronic degrees of freedom, to classical force fields, to effective coarse-grained models, up to continuum models on the macroscopic scale. Limited computational capacities then imply that these levels correspond to increasing length and time scales that can be accessed at the expense of decreasing molecular detail. Indeed, for many applications details of the molecular interactions are irrelevant and models can be devised based on symmetry considerations and conservation laws alone, with the Navier-Stokes equations being a prominent example for a continuum model of fluid dynamics.

On intermediate levels, systematic approaches to structural coarse-graining have been developed~\cite{mull02,pete09} such as iterative Boltzmann inversion, which determines pair potentials for a reduced set of degrees of freedom from (partial) radial distribution functions obtained either from experiments or atomistic simulations. These relevant degrees of freedom have to be provided by the user and are typically informed by physical or chemical insights. One side effect of such a procedure is the loss of dynamical information, which is understood qualitatively (removed degrees of freedom act as a ``bath'' leading to stochastic dynamics~\cite{oett07}) but still hard to control.

Quite a different approach is followed through the construction of Markov state models~\cite{chod07,noe08,prin11}, which aim to bridge the long time scales involved in, \emph{e.g.}, the folding of proteins from an initially disordered coil to the native state. These models are built by clustering microstates into a few mesoscopic states that are kinetically distinct, \emph{i.e.}, they correspond to basins that are separated by free energy barriers with a time-scale separation between fast intra-basin transitions and slow inter-basin transitions. In the case of a complete separation this can be exploited to endow the mesoscopic states with a Markovian dynamics for the slow transitions assuming quasi-equilibrium for the fast transitions.

One should note that a rigorous approach to dynamically consistent coarse-graining has been around for quite some time using projection techniques~\cite{grabert,zwanzig}. By projecting the dynamics onto the subspace spanned by the reduced variables, a non-Markovian generalized Langevin equation for the time evolution in this subspace is obtained, which is formally exact but now takes on the form of an integro-differential equation. The complexity of the original problem is, therefore, preserved as ones trades the reduction of the degrees of freedom for a memory kernel. Further approximations are needed to obtain a numerically tractable problem. In the case of a clear separation of time scales one can employ the Markovian approximation~\cite{hijo10}. But even then the resulting equations are highly non-trivial both from a physical and mathematical point of view. Within this framework attempts have been made to derive dynamical density functional theory~\cite{espa09} and its extension to non-equilibrium situations~\cite{witt13,schm13a}.

Indeed, another layer of complexity is added when going away from thermal equilibrium to driven systems. \emph{Stochastic thermodynamics} has emerged as a comprehensive theoretical framework in particular for driven systems that are still in contact with a heat reservoir that itself remains in equilibrium~\cite{seif12}. Here two classes of non-equilibrium situations can be distinguished: time-dependent processes with dynamics that still obey detailed balance and non-equilibrium steady states with dynamics that explicitly break detailed balance through non-conservative forces and/or flows~\cite{spec08}. Fluctuating path functionals like work, heat, and entropy production obey certain symmetries called fluctuation theorems~\cite{seif12}, which have been studied for (chemical) networks of discrete states~\cite{andr06,andr07a,schm07}. Hidden (unobservable) degrees of freedom modify the fluctuation theorems in various ways~\cite{mehl12,leon13}, and the consequences of coarse-graining have been discussed for removing fast states~\cite{pugl10}, ``bridge'' states~\cite{alta12}, and the clustering of microstates~\cite{raha07,espo12}.

So far, only relatively simple models have been studied appealing to stochastic thermodynamics including a network model of kinesin with six states~\cite{alta12,alta12a} and a model for motor proteins~\cite{zimm12,zimm15}. Building on these works, the purpose of the present paper is to extend the Markov state model approach to driven many-body systems and to construct coarse-grained models that are consistent with stochastic thermodynamics. The paper is organized as follows: in section~\ref{sec:back} we briefly revisit the prerequisites, in particular how to obtain a network of discrete states from atomistic simulation data and introduce the relevant notions from stochastic thermodynamics. In particular \emph{cycles} and the decomposition of networks into cycles~\cite{schnakenberg1976network,hill,zia07,alta12a,pole15} will play a crucial role. The actual algorithm consists of two parts: the identification and clustering of cycles discussed in section~\ref{sec:cc} followed by the actual coarse-graining \emph{in cycle space} described in section~\ref{sec:cg}. Finally, we provide some critical remarks before concluding.

\section{Motivation and basic idea}

In order to maintain a non-equilibrium steady state, work has to be supplied constantly, which is dissipated as heat $q$ into the environment. This non-vanishing entropy production allows for the \emph{transport} of some quantity $X$, which might be the distance travelled by a particle, the number of molecules produced in a chemical reaction, etc. Although the mapping to Markovian dynamics proposed in this paper is not rigorous, there is evidence that the preservation of entropy production also implies the preservation of the major dynamical properties as well as macroscopic transport. Indeed, very recently Barato and Seifert have conjectured that the dissipation of a process leading to a squared relative uncertainty $(\delta X)^2\equiv\mean{(X-\mean{X})^2}/\mean{X}^2$ is at least
\begin{equation}
  \label{eq:diss:bound}
  \mean{q} = \mean{\dot q}t \geq \frac{2}{(\delta X)^2},
\end{equation}
where $t$ is the duration of the process~\cite{bara15}. We employ dimensionless quantities throughout. In particular, entropies are measured in units of Boltzmann's constant $k_\text{B}$ and energies in units of $k_\text{B}T$, where $T$ is the temperature of the surrounding heat bath.

The first step is to construct a discrete Markov state model and to determine the transition rates from particle-resolved simulation data. Although already this step is non-trivial it is not the focus of the present manuscript, in which we study a sufficiently simple model to employ a straightforward construction. For our purposes it is important to include many discrete states to approximate the entropy production of the particle-resolved model as closely as possible. Entropy production is related to cycles in the network of these discrete states. It seems quite clear that removing states will lead to ``cutting open'' some of these cycles and thus to a reduced entropy production~\cite{pugl10}. Our approach to this challenge is an additional step before the actual coarse-graining, which consists of identifying communities of cycles with similiar properties. The crucial step is then to identify a representative for each community, which will receive the entire entropy production of its community. All other cycles are then removed, and with them all states that are not visited by one of the representatives. This procedure preserves the entropy production rate, and Eq.~\eqref{eq:diss:bound} therefore implies that fluctuations of (measurable) transported quantities obey the same lower bound in the original and the coarse-grained system.

\section{Background}
\label{sec:back}

\subsection{Model system: driven particle in a double well potential}
\label{sec:modelsystem}

We will illustrate our approach to coarse-graining with a specific example: a Brownian particle in a two-dimensional symmetric potential driven by a non-conservative force $\vect{F}_{\text{nc}}$. Working in two dimensions allows to directly visualize the configurations space as well as cycles, although the discussed algorithm is more generally applicable to many-body systems.

\begin{figure}
  \hfill\includegraphics[width=.8\linewidth]{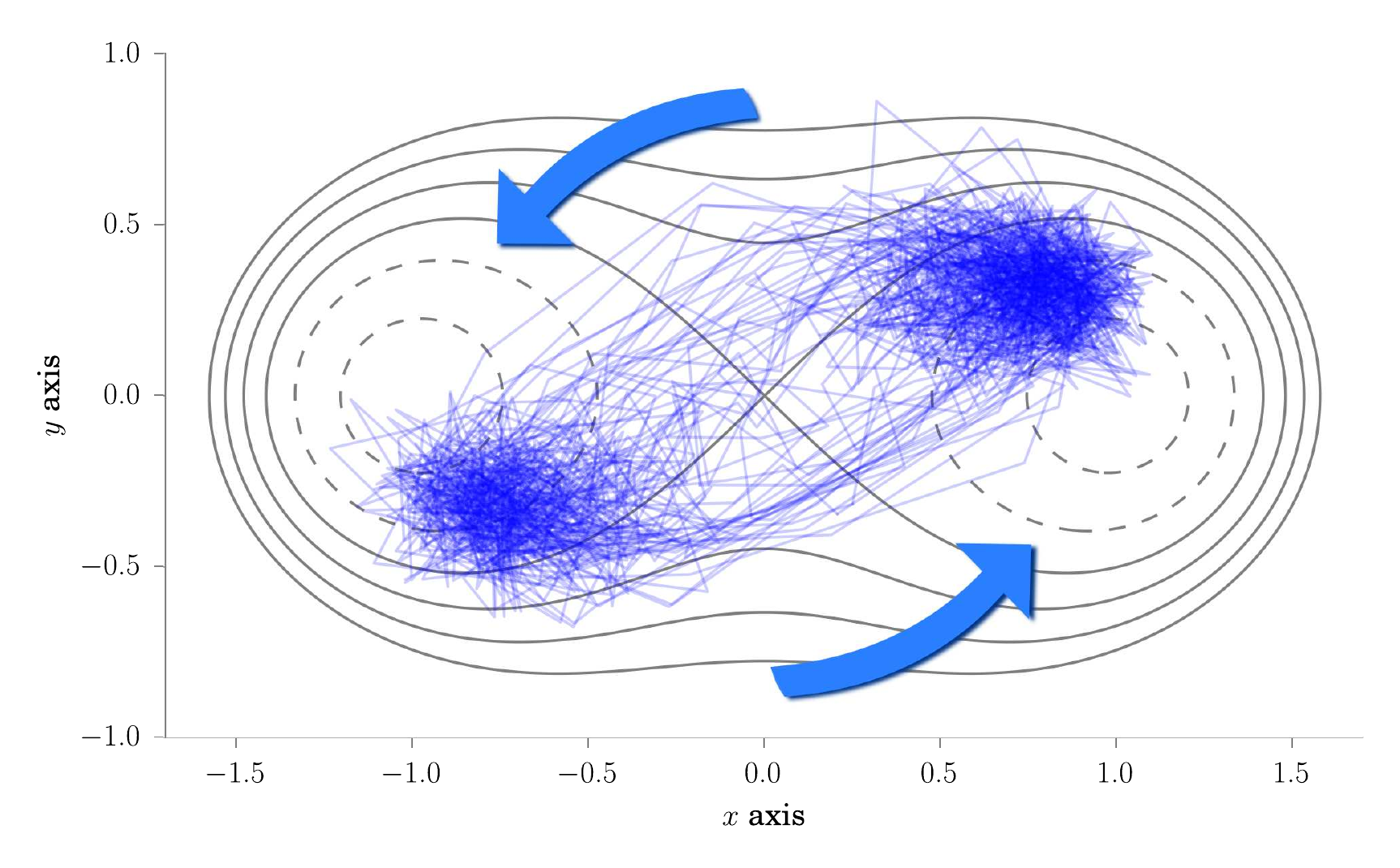}
  \caption{Contour lines of the symmetric two-dimensional potential eq.~\eqref{eq:pot} and exemplary trajectory in the presence of the non-conservative force (the arrows indicate the direction of the driving). The system is still invariant under inversion with the loci of highest probability being shifted from the minima of the potential.\label{fig:traj}}
\end{figure}

The equation of motion follows as
\begin{equation}
  \label{eq:bd}
  \dot{\vect{x}} = -\nabla U + \vect{F}_{\text{nc}} + \vect{\eta}(t),
\end{equation}
where $\vect{\eta}(t)$ is a random force with correlations $\langle \vect\eta(t)\vect{\eta}(t')\rangle = 2\delta(t-t')$ and
\begin{equation}
  \label{eq:pot}
  U(\vect{x}) = \frac{x^4}{4} + \frac{1}{2}(y^2 - x^2 + x^2y^2)
\end{equation}
is the potential energy. As non-conservative force we choose $\vect{F}_{\text{nc}}(\vect{x}) = \omega\,(-y,x)^T$, where $\omega$ denotes the driving strength. Figure~\ref{fig:traj} shows the contour lines of the potential and an exemplary trajectory. Due to $\vect{F}_{\text{nc}}$, the particle trajectory does not obey the symmetry of the conservative potential anymore.

\subsection{Building Markov state models, in a nutshell}

We start by giving a short overview of how to create a Markov state model from, in general, molecular dynamics simulations, while here we employ Brownian dynamics (BD) propagating eq.~\eqref{eq:bd}. A detailed review can be found in~\cite{prinz2011markov}. The first step in building a Markov state model is to approximate the continuous state space (only the positions are taken into account) by a set of discrete states that we will refer to as the microstates. The spatial discretization is typically obtained by cluster analyses, e.g. via the $k$-means algorithm. After a proper spatial discretization is found, the continuous $d$-dimensional state space is discretized into $k$ partitions. Each of the $k$ partitions is represented by its center, called \emph{centroid}, allowing the original data points to be mapped onto the centroids by minimizing their Euclidean distance.

Once the full state space is discretized, the BD trajectory can be projected onto its centroids, storing its dynamical information as a simple sequence of centroid indices. The dynamical information can be extracted from this sequence of centroids by counting the number of transitions $C_{i\rightarrow j}$. Whenever centroid $i$ is followed by $j$ we increase $C_{i\rightarrow j}$ by one. Finally, the transition matrix $T$ is approximated by
\begin{equation}
  T_{i\rightarrow j} = \frac{C_{i\rightarrow j}}{\sum_j C_{i\rightarrow j}},
\end{equation}
which is also the maximum probability estimator for the true transition operator~\cite{prinz2011markov}. If $\Delta t$ is chosen small enough, we can further expand
\begin{equation}
  T = \exp{(\Delta t W)} \approx \id + \Delta t W + O(\Delta t^2)
  \label{eq:transition_matrix}
\end{equation}
with \emph{rate matrix} $W$. It is important to note that for driven systems neither $W$ nor $T$ are symmetric in general. Hence, their eigenvalues are complex, making the identification of metastable sets difficult. The reason is that the real part of the spectrum does not necessarily offer a gap, an illustrative example for which can be found in ref.~\citenum{conrad2014modularity}.

\subsection{Stochastic thermodynamics}

\subsubsection{Steady state Markov models}

The discrete states and their corresponding transition rates can be represented by a graph $G(V,E)$, where vertices $V$ represent the states and edges $E$ the transitions connecting the states. The number of vertices in $G$ is denoted by $|V|$, while the number of edges is $|E|$.

If all transition rates $w^i_j=W_{i\ra j}$ are known\footnote{We adopt the notation that edges are written as upper (source) and lower (destination) indices, \emph{i.e.}, $X^i_j$ is the value of $X$ taken in the transition $i\ra j$.}, the time evolution of the state probabilities is given by the master equation
\begin{equation}
  \label{eq:master:I}
  \frac{\partial p_j(t)}{\partial t} = 
  \sum_{i\neq j}w^i_jp_i(t)-w^j_ip_j(t) = \sum_{i\neq j}\Phi^i_j(t) - \Phi^j_i(t)
\end{equation}
with normalization
\begin{equation}
  \sum_i p_i(t) = 1.
\end{equation}
Here $\Phi^i_j\equiv w^i_jp_i$ are the directional probability fluxes indicating the amount of probability flowing from state $i$ to state $j$ per unit time. The subsequent description requires that the graph spanned by the Markov model is (i)~ergodic, \emph{i.e.}, every state can be reached by every other state in finite time, and (ii)~the transition rates are reversible, \emph{i.e.}, if $w^i_j>0$ then also $w^j_i>0$. Hence, there is one edge for the forward and one edge for the backward transition between any two states that are connected through non-zero transition rates.

If the system is in a steady state the probability distribution, and hence probability fluxes, are time independent and we can drop the time argument, $p_i(t)=p_i$. The left hand side of equation~\eqref{eq:master:I} becomes zero,
\begin{equation}
  \label{eq:master:II}
 \frac{\partial p_j}{\partial t} = 0 
 = \sum_{i\neq j}\Phi^{i}_j - \Phi^{j}_i.
\end{equation}
In the following we assume that the system resides in a (non-equilibrium) steady state.

A special steady state is identified as thermodynamic equilibrium. Here each summand of the right hand side of eq.~\eqref{eq:master:II} vanishes individually, i.e. $\Phi^{i,\text{eq}}_j - \Phi^{j,\text{eq}}_i = 0$, while the 
condition $w^{i}_j p^\text{eq}_i = w^{j}_i p^\text{eq}_j$ is called detailed-balance stating that the amount of transported probability per unit time in the transition $i\rightarrow j$ equals the amount of the reverse transition $j\rightarrow i$.
If detailed-balance is broken a net probability current flows from $i\rightarrow j$ and eq.~\eqref{eq:master:II} can be identified as Kirchhoff's current law stating that the same amount of probability flowing into state $i$ is also flowing out.

\subsubsection{Entropy production}

Following refs.~\citenum{schnakenberg1976network,seifert2005entropy}, the total average entropy production rate $\langle\dot{S}_{\text{tot}}\rangle$ reads
\begin{equation}
  \label{eq:entropy_production}
  \langle\dot{S}_{\text{tot}}\rangle 
  = \sum_{ij}\Phi^i_j\ln{\left(\frac{\Phi^i_j}{\Phi^j_i}\right)}
  = \underbrace{\sum_{ij}\Phi^i_j 
    \ln{\left(\frac{p_i}{p_j}\right)}}_{\langle\dot{S}_{\text{sys}}\rangle} 
  +\underbrace{\sum_{ij}\Phi^i_j
    \ln{\left(\frac{w^i_j}{w^j_i}\right)}}_{\langle\dot{S}_{\text{med}}\rangle}.
\end{equation}
The first term is identified as the time derivative of the Gibbs-entropy $S_{\text{sys}}\equiv-\sum_ip_i\ln{p_i}$ and thus describes the entropy change of the system itself. The second term represents the coupling of the system with its environment (medium) in such a way that the system cannot reach equilibrium.  The coupling can be, for example, caused by a heat or particle flux flowing from the medium into the system. Only the sum of both entropy production rates is larger than zero in accordance with the second law. If the system is in a non-equilibrium steady state (NESS), the first term vanishes and consequently $\langle\dot{S}_{\text{tot}}\rangle = \langle \dot{S}_{\text{med}}\rangle$ balance each other.

A second important quantity related to entropy production are \emph{affinities}, or generalized thermodynamic forces~\cite{schnakenberg1976network}. The edge affinity between two states is defined as
\begin{equation}
  \label{eq:edge_affinity}
  A^i_j \equiv \ln{\left(\frac{\Phi^i_j}{\Phi^j_i}\right)} = \underbrace{\ln{\left(\frac{w^i_j}{w^j_i}\right)}}_{\sigma^i_j} +\ln{\left(\frac{p_i}{p_j}\right)},
\end{equation}
which can, again, be split into two parts. The first part $\sigma^i_j$ is identified as the entropy produced in the medium~\cite{seifert2005entropy} for every transition $i\ra j$. Interestingly, neither the affinities nor the dissipated heat depend on the time scales governing the dynamics of the system as in both expressions only the ratio of rates are taken into account.

Finally, the central concept in our approach to coarse-graining is \emph{cycles}. A cycle is an ordered set of states (vertices), at the end of which the starting state is reached again. Cycles that differ only in their cyclic permutation of vertices are considered identical. For example, $\{1,2,3,1\} =\{2,3,1,2\} =\{3,1,2,3\}$ all denote the same cycle but $\{2,1,3,2\}$ is a different cycle. We further distinguish between \emph{trivial cycles} (cycles with only two different states, \emph{i.e.}, $\{i,j,i\}$), and \emph{non-trivial cycles} that contain at least three different states.

Two types of observables\footnote{Throughout, greek indices denote cycles.} can be distinguished for each cycle:
\begin{itemize}
\item[(i)] Current-like observables that are summed along the edges corresponding to each cycle. Consider an observable defined on the edges of a graph, that is given by a matrix $X\in\mathbb{R}^{|V|\times |V|}$. The cycle observable is computed as
  \begin{equation}
    \label{eq:def_c_av_x}
    X_\alpha = \sum_{\ij\in \alpha} X^i_j.
  \end{equation}
  The notation $\sum_{\ij\in\alpha}$ denotes the summation over all directed edges $i\rightarrow j$ that are part of cycle $\alpha$.
\item[(ii)] State-like observables that are summed over the states forming a cycle. Here the observables are defined on the graph vertices given by a vector $Y\in\mathbb{R}^{|V|}$, and thus
  \begin{equation}
    \label{eq:def_c_av_y}
    Y_\alpha = \sum_{i\in \alpha} Y_i.
  \end{equation}
\end{itemize}
An example for the latter is given by the average cycle period $\tau_\alpha=\sum_{i\in\alpha}\tau_i$ (the average time that is spend in cycle $\alpha$) with $\tau_i=\left(\sum_jw^i_j\right)^{-1}$ being the average time spend in state $i$.

\subsubsection{Cycle affinities}

The most prominent example for a current-like observable is the \emph{cycle affinity}
\begin{equation}
  A_\alpha = \sum_{\ij\in \alpha} A^i_j.
\end{equation}
It has two important properties:
\begin{itemize}
\item[(i)] The cycle affinity of trivial cycles is always zero since affinities are anti-symmetric, $A^i_j = - A^j_i$, and thus $A_{\{i,j,i\}} = A^i_j + A^j_i = A^i_j - A^i_j = 0$.
\item[(ii)] All cycle affinities are independent of the state probabilities and thus can be expressed by $\sigma^i_j$'s only.
\end{itemize}
To prove (ii),
\begin{eqnarray}
  \nonumber
  A_{\{i,j,...,n,i\}} = A^i_j + \dots + A^n_i &= \ln{\left(\frac{w^i_jp_i\dots w^n_ip_n}{w^j_ip_j\dots w^i_n p_i}\right)} \\
  \label{eq:cycle_affinity}
  &= \ln{\left(\frac{w^i_j\dots w^n_i}{w^j_i\dots w^i_n}\right)} = \sigma^i_j + \dots +\sigma^n_i.
\end{eqnarray}
Since the $p_i$'s are state functions, going along a cycle they cancel pairwise. A remarkable property of a NESS is thus that the average entropy production can be calculated either by using the $A^i_j$'s or $\sigma^i_j$'s, which correspond to $S_{\text{tot}}$ and $S_{\text{med}}$, respectively.

\section{Cycle clustering}
\label{sec:cc}

\subsection{Cycle-flux decomposition}

For convenience, we define the indicator
\begin{equation}
  \chi^i_\alpha = \cases{1&if vertex $i$ is in cycle $\alpha$\\0&otherwise}
\end{equation}
and passage function
\begin{equation}
  \chi^i_{j,\alpha} = \cases{1&if directed edge $i\ra j$ is in cycle $\alpha$\\0&otherwise,}
\end{equation}
depending on whether a vertex $i$ or directed edge $i\ra j$ is part of a cycle $\alpha$, respectively. Following the ideas of refs.~\citenum{macqueen1981circuit,kalpazidoubook,alta12a}, the cycle-flux decomposition expresses the flux field $\Phi$ through a linear combination of cycles,
\begin{equation}
  \label{eq:decomp}
  \Phi^i_j = \sum_\alpha \varphi_\alpha \chi^i_{j,\alpha}.
\end{equation} 
The \emph{cycle weight} $\varphi_\alpha$ quantifies how much probability flows through cycle $\alpha$ per unit time. As shown in refs.~\citenum{macqueen1981circuit,kalpazidoubook}, $\varphi_\alpha\geq0$ is non-negative and the decomposition eq.~\eqref{eq:decomp} always exists if $\Phi$ satisfies the steady state condition (Kirchhoff's current law). However, this decomposition is not unique as discussed in the next section~\ref{sec:algo}.

Using the cycle observables eq.~\eqref{eq:def_c_av_x} and inserting the cycle decomposition eq.~\eqref{eq:decomp}, the average of a current-like observable
\begin{equation}
  \mean{X} = \sum_{ij} \Phi^i_jX^i_j = \sum_\al \vhi_\al \sum_{ij}X^i_j\chi^i_{j,\al} 
  = \sum_\al \vhi_\al X_\al
\end{equation}
can be expressed as a cycle average with weights $\vhi_\al$~\cite{alta12a}. In particular, the cycle average of affinities
\begin{equation}
  \mean{A} = \sum_\al \vhi_\al A_\al = \sum_{ij} \Phi^i_jA^i_j 
  = \mean{\dot S_\text{tot}}
\end{equation}
equals the total average entropy production with the definition of edge affinities eq.~\eqref{eq:edge_affinity}. Furthermore, we define the conditional average entropy production of a single cycle $s_\alpha\equiv\vhi_\alpha A_\alpha$ so that the total average entropy production becomes the sum $\mean{\dot S_\text{tot}}=\sum_\alpha s_\alpha$.

\subsection{Algorithm}
\label{sec:algo}

Several types of algorithms have been proposed in the literature to accomplish the decomposition eq.~\eqref{eq:decomp}~\cite{schnakenberg1976network ,quianbook,macqueen1981circuit}. For example the ``method of derived chain'' introduced in ref.~\citenum{quianbook}, which is stochastic in nature and has the advantage that the cycle weights $\vhi_\al$ in eq.~\eqref{eq:decomp} are unique and that they correspond to the mean number of passages through cycle $\al$. However, negative cycle entropies ($s_\al<0$) may occur, which greatly complicates the identification of communities (as will become clear shortly). Moreover, the number of cycles used in the decomposition can be orders of magnitude larger than for the other approaches discussed below. Another important type of cycle decompositions, first mentioned by Schnakenberg~\cite{schnakenberg1976network}, considers fundamental cycles that span a basis of the cycle space. Although the number of contributing cycles is as small as for the described algorithm below, the fundamental cycles are not unique and can have negative cycle entropies as well.

We employ a variant of the ``cycle-flux'' decomposition introduced by MacQueen~\cite{macqueen1981circuit}. This is a deterministic algorithm that has a polynomial complexity in the number of vertices $|V|$, making it computationally affordable even for large graphs. Again, the decomposition (and thus the cycle weights $\vhi_\al$) is not unique but rather depends on the initial cycle sequencing, where already a minor variation in the sequencing can lead to different cycle weights. In particular, some cycle weights become zero while others become non-zero. An example illustrating this arbitrariness is given in ref.~\citenum{alta12a}, where the cycle-flux decomposition is applied to the totally asymmetric simple exclusion process (TASEP) model.

Let $C=\{\al\}$ denote the set of \emph{all} possible cycles. Only the subset $C'\subset C$ of cycles with non-vanishing cycle weights contribute to averages, an upper bound $|C'|\leq B$ of which is given by the Betti number $B\equiv |E|-|V|+1$~\cite{kalpazidoubook}. The problem is that one cannot know beforehand what the contributing cycles are, meaning that theoretically all possible cycles are needed to ensure a complete decomposition, although most of them will have zero weights especially for graphs with many vertices. Already the number of possible cycles of an undirected graph is bounded by $B\leq|C|\leq 2^B$~\cite{volkmann_graph_theory}.

We now describe the algorithm in more detail. To this end, we sort cycles,
\begin{equation}
  C' = \{\underbrace{\al_0,\al_1,...,\al_{|E|/2}}_{\text{trivial cycles}},
  \dots,\al_{|C|}\},
\end{equation}
where first come the trivial cycles followed by the non-trivial cycles containing three or more states. The number of trivial cycles is given by half the number of edges, $|E|/2$. Initialize $\tilde\Phi\leftarrow\Phi$ and take the first cycle $\al$ from $C'$:
\begin{enumerate}
\item Take all fluxes $\tilde\Phi$ along cycle $\al$ and find the smallest value, which becomes the cycle weight $\vhi_\al$,
  \begin{equation}
    \vhi_\al = \min_{\ij\in\al}\{\tilde\Phi^i_j\}.
  \end{equation}
\item Subtract $\vhi_\al$ from all fluxes along $\al$, $\tilde\Phi^i_j\ra\tilde\Phi^i_j-\vhi_\al\chi^i_{j,\al}$. The new flux field has at least one edge less.
\item Advance to the next cycle $\al$ in $C'$ and repeat.
\item The algorithm stops when the residuum $||\tilde\Phi||_{\text{max}}$ has become smaller than some threshold. 
\end{enumerate}
After the first $|E|/2$ iterations, all trivial cycles are subtracted and hence the flux field $\tilde\Phi$ at this stage contains only irreversible edges, \emph{i.e.}, if $\tilde\Phi^i_j>0$ then $\tilde\Phi^j_i=0$. Hence, the fluxes have become the (positive) currents of the original graph, $\Phi^i_j-\Phi^j_i\ra\tilde\Phi^i_j$, where $\tilde\Phi^i_j>0$ since the algorithm subtracts the smaller value. The original affinities $A^i_j$ along the remaining edges are thus positive, cf. eq.~\eqref{eq:edge_affinity}, and hence all cycle affinities for the remaining non-trivial cycles are positive. An illustrative example for a concrete graph is presented in appendix~\ref{sec:cycle:decomp}. Note that this implies $s_\al>0$ for all non-trivial cycles.

In appendix~\ref{sec:cycle:det} an algorithm is outlined how to obtain all contributing cycles $C'$. Removing the trivial cycles, we can thus further reduce the set of cycles $\Cn\subset C'$ to be considered in the following with $|\Cn|\leq B-|E|/2$.

\subsection{Cycle clustering}

\subsubsection{Absence of dominant cycles}

\begin{figure}
  \hfill\includegraphics[width=.8\linewidth]{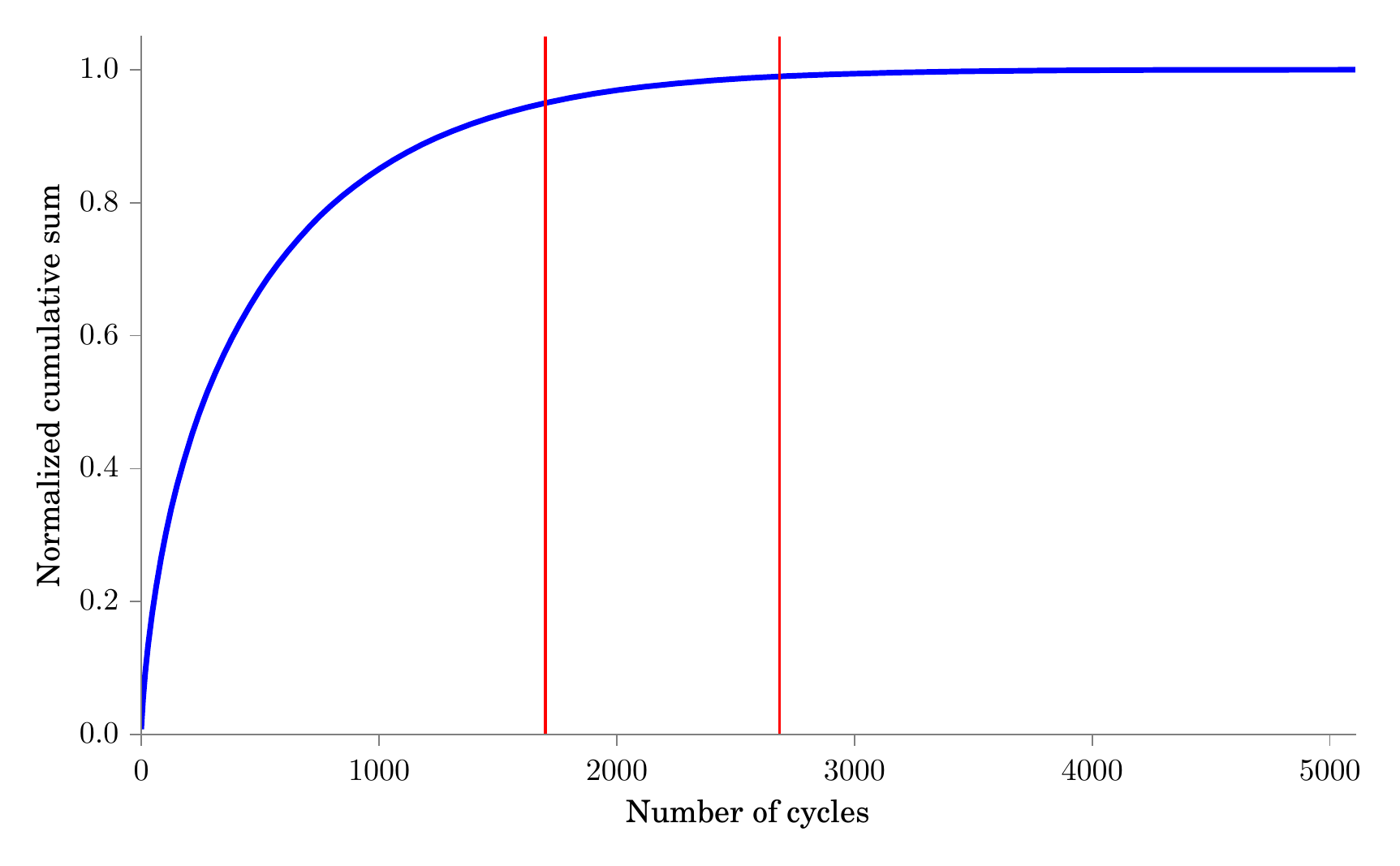}
  \caption{Normalized cumulative sum eq.~(\ref{eq:cumsum}) of cycle entropy production of our model system. The first red line marks the number of cycles to recover 95\% and the second to recover 99\% of the total entropy production.\label{fig:cumsum}}
\end{figure}

Markov state models that are constructed from molecular dynamics simulations contain thousands, or even millions, of cycles. This is in stark contrast to the semi-empirical models discussed so far containing a handful of -- already coarse-grained -- states~\cite{alta12,zimm12}. At this stage it is important to realize that there are \emph{no dominant cycles} in terms of the entropy production. This is illustrated in figure~\ref{fig:cumsum} for our specific model system. We have chosen $|V|=200$ (this is an input to the $k$-means clustering algorithm) and found $|E|=10612$ edges implying $B=10413$ with non-trivial cycles $|\Cn|\leq5107$. Shown is the normalized cumulative sum
\begin{equation}
  \label{eq:cumsum}
  \tilde s_n \equiv \frac{\sum_{\al\leq n}s_\al}{\sum_{\al}s_\al}
\end{equation}
of the first $n$ cycles for sorted cycle entropies $s_0\geq s_1\geq\dots$ Already for our simple model system a large number of cycles contribute and no dominant cycles appear. To recover 95\% of the total entropy production, almost 2000 cycles have to be included, although many of the non-trivial cycles do not contribute substantially to the overall entropy production.

\subsubsection{Linking cycles}

Our approach in the following is based on the idea that many cycles are similar in their length, traversed states (region in state phase), and cycle affinities. We will propose how to quantify this ``similarity'' of cycles and how to group these cycles together, forming \emph{cycle communities}.

The cycles in $\Cn$ can again be interpreted as vertices of a graph $\mathcal G=(\Cn,\Omega)$. It is important to realize that we have a considerable freedom in defining edge weights (analogous to the original transition rates $w^i_j$). Here we opt to link cycles $\al$ and $\beta$ using
\begin{equation}
  \Omega_{\alpha\beta} \equiv \frac{2 \sum_i p_i \chi_{i,\alpha}\chi_{i,\beta}}
  {\sum_ip_i\chi_{i,\alpha}+\sum_ip_i \chi_{i,\beta}},
\end{equation}
which is symmetric as emphasized by the subscripts. Two cycles are linked only if they share at least one vertex on the original graph, while the connectivity strength $0\leq\Omega_{\al\beta}\leq 1$ depends on the cumulated probability of the shared states. Other proposals for linking cycles can be found in refs.~\citenum{alta12a,conrad2014modularity}.

\begin{figure}
  \centering
  \includegraphics[width=0.5\linewidth]{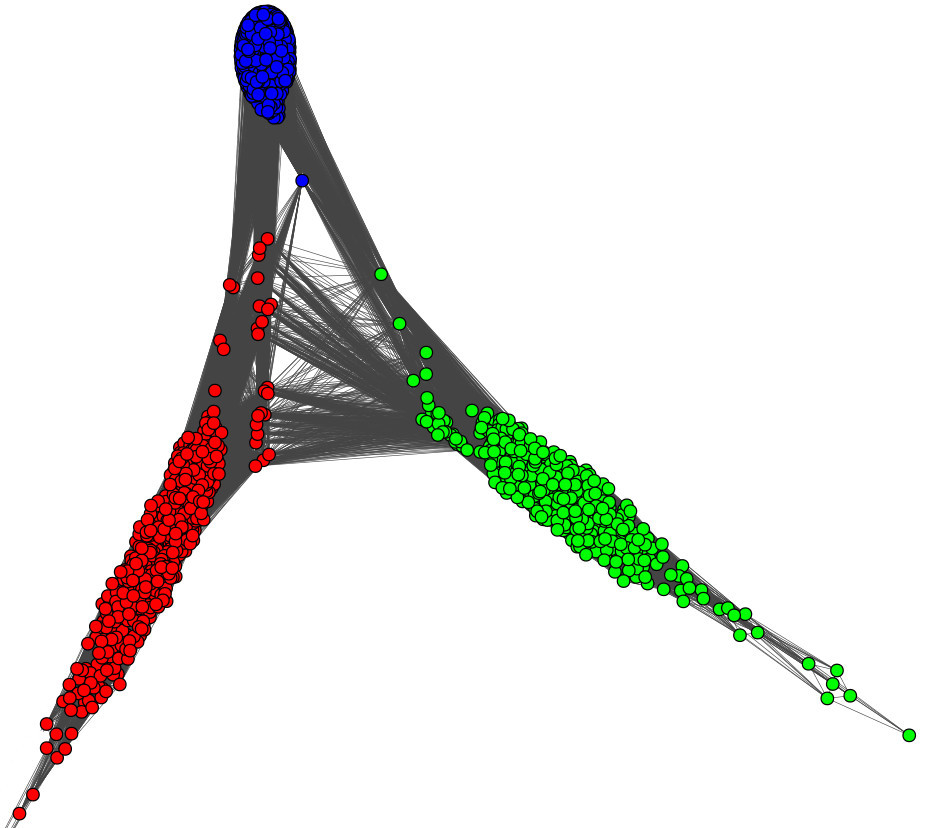}
  \caption{Cycle graph constructed from our model system. Each vertex represents a non-trivial cycle. Colors indicate cycles belonging to the same cycle community. The vertices are arranged utilizing a force-directed graph drawing algorithm.\label{fig:cycle_space}}
\end{figure}

In practice, we found that the resulting graph is connected too strongly and that better results are achieved if $\Omega$ is modified using additional information. We thus cut a link if (i)~two cycles have dissimilar cycle affinities and (ii)~if their spatial extensions differ too much. For the later, the spatial extension of a cycle is defined as the maximal Euclidian distance of centroid positions
\begin{equation}
  \dm_\al \equiv \max_{(i,j)\in\alpha} ||\vect x_i - \vect x_j||_2,
\end{equation}
where $\vect x_i$ is the state-space vector representing centroid $i$. Hence,
\begin{eqnarray}
  \label{eq:connectivity}
  \Omega_{\alpha\beta} \leftarrow \cases{0&if $\frac{\min\{A_\alpha,A_\beta\}}{\max\{A_\alpha,A_\beta\}} \le \xi_a$\\0&if $\frac{\min\{\dm_\alpha,\dm_\beta\}}{\max\{\dm_\alpha,\dm_\beta\}} \le \xi_d$\\\Omega_{\alpha\beta}&otherwise}
\end{eqnarray}
with $\xi_a,\xi_d\in(0,1)$. Specifically, for our model system we have set $\xi_a=\xi_d=\frac{1}{2}$. The resulting cycle graph is shown in figure~\ref{fig:cycle_space}.

\subsubsection{Cycle communities}

Having defined $\Omega$ to be symmetric and thus undirected, we can make use of well-developed community finding algorithms for undirected graphs. Formally, the communities are disjoint sets $C_l$ of the non-trivial cycles,
\begin{equation}
  \Cn = \bigcup_l C_l.
\end{equation}
A community is a group of vertices that share similar properties and are consequently stronger connected (higher link density) to vertices inside their community than with vertices from other communities. The number of communities is inherent to its graph and not preassigned. A review of the most important algorithms can be found in ref.~\citenum{fortunato2010community}. In general, a measure of how good community finding algorithms perform is the modularity function $Q$, which measures the link density inside communities as compared to other communities by assigning it a value between -1 and 1~\cite{newman2004fast}. It is defined as 
\begin{equation}
  \fl
  \label{eq:modularity_function}
  Q \equiv \frac{1}{2m}\sum_{\alpha¸\beta}\left(\Omega_{\alpha¸\beta} - \frac{k_\alpha k_\beta}{2m}\right)\times\cases{1&if $\alpha$ and $\beta$ belong to same community\\0&otherwise}
\end{equation}
with $k_\beta\equiv\sum_\alpha \Omega_{\alpha\beta}$ and $m\equiv\frac{1}{2}\sum_\beta k_\beta$.

\begin{figure}
  \hfill\includegraphics[width=.8\linewidth]{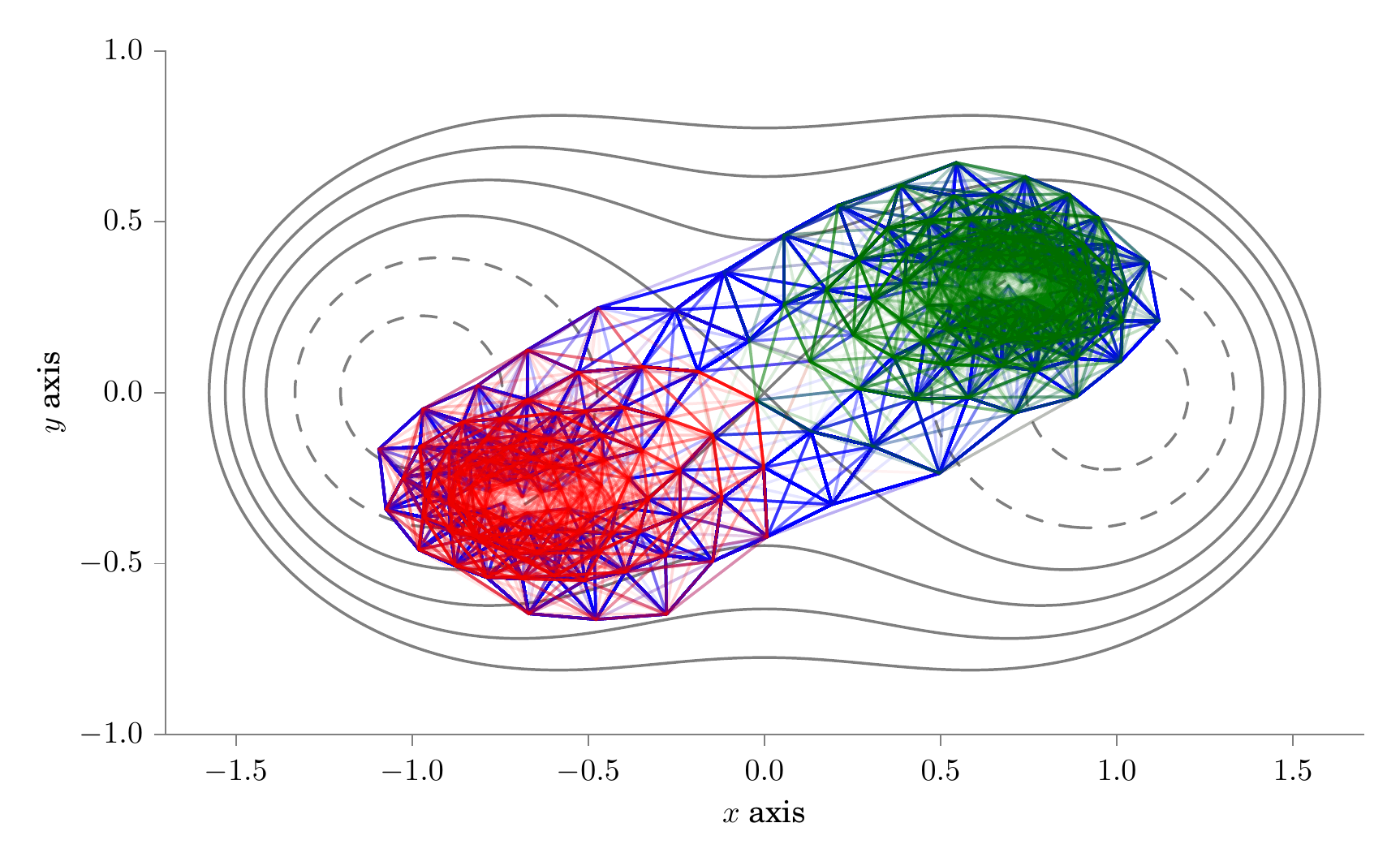}
  \caption{Cycle communities in state space, where edges are colored according to the community of their cycles, cf. figure~\ref{fig:cycle_space}. Three communities with red, green, and blue have been identified by the community finding algorithm.\label{fig:cycle_communities}}
\end{figure}

Specifically, we use a community finding algorithm~\cite{blondel2008fast} that has been shown to be fast and achieves good results also for a large number of cycles (vertices in the cycle space). Here we are interested in finding communities among all non-trivial cycles that were found by applying the cycle-flux decomposition algorithm described in the previous section~\ref{sec:algo}. The communities found for our model system are displayed in figure~\ref{fig:cycle_space} in cycle space and in figure~\ref{fig:cycle_communities} for the original state space. Three communities have been detected: The red and green colored communities are in agreement with the loci of highest probability, cf. figure~\ref{fig:traj}, while the blue community contains cycles connecting these two loci. This result agrees well with the intuitive picture of entropy production due to the interplay between conservative and non-conservative forces for the particle trapped in a basin, with rare transitions between both basins.

\section{Coarse graining}
\label{sec:cg}

\subsection{Cycle representatives}

We have now identified communities of cycles with similiar properties determined through the link strength $\Omega_{\al\beta}$. However, it is not possible to compute something like an average cycle because the cycle space does not have any physical metric. Furthermore, we are restricted to the original states and transition rates (or rather $\sigma^i_j$'s) as they have a physical meaning, \emph{i.e.}, position in state space and medium entropy production per transition, respectively. To overcome this problem, we determine one cycle out of each cycle community, which we will refer to as the (community) \emph{representative}. This cycle is then rescaled in order to preserve the entropy production of the whole community, and the other cycles are discarded.

To this end, we rewrite the modularity function eq.~\eqref{eq:modularity_function} as the sum over all community modularities
\begin{equation}
  Q = \sum_l Q_l = \sum_l\frac{1}{2m} \left(\sum_{\alpha,\beta} 
    \Omega^{(l)}_{\alpha\beta} - \frac{k^{(l)}_\alpha k^{(l)}_\beta}{2m}\right),
\end{equation}
where $l$ denotes the $l$-th community. Our task is to find the cycle $\gam$ for each cycle community $l$ that maximizes $Q_l-\Qr$, with $\Qr$ being the modularity of the $l$-th community \emph{without} cycle $\gam$. In other words, we want to identify the cycle $\gam$ for each community that has the biggest impact on the community modularity. In particular, $\Qr$ increases if cycle $\gam$ matches poorly and decreases when it provides many links to other cycles inside its community. After some algebra, the difference is given by
\begin{equation}
  \label{eq:representative}
  Q_l - \Qr = \frac{k^{(l)}_\gam}{m}\left(1+\frac{k^{(l)}_\gam - 
      \sum_\alpha k^{(l)}_\alpha}{m}\right). 
\end{equation}
For our model system, the three representatives found from maximizing the modularity difference are illustrated in figure~\ref{fig:cycle_representative}.

\subsection{Rescaling}

After determining the representatives, the original fluxes $\Phi^i_j$ and transition rates $w^i_j$ have to be rescaled. The physical constraints are:
\begin{enumerate}
\item The total entropy production $\mean{\dot S_\text{tot}}$ is preserved.
\item The $\sigma^i_j$'s of all non-vanishing edges are preserved.
\item The cycle affinities $A_\alpha$ of all contributing cycles are preserved.
\end{enumerate}
It is easy to check that (iii) is valid if (ii) is fulfilled, cf. eq.~\eqref{eq:cycle_affinity}. For every cycle community, we define a \emph{community} entropy production
\begin{equation}
  S_l \equiv \sum_{\al\in C_l} s_\al \quad\text{with}\quad
  \mean{\dot S_{\text{tot}}} = \sum_lS_l.
\end{equation}
Because all non-trivial, and thus possibly entropy producing, cycles are partitioned into communities, the total entropy production of the original graph is the sum of all community entropy productions. In what follows, all rescaled quantities are labeled by hats, \emph{e.g.}, $\vhi_\alpha\ra\hat{\vhi}_\alpha$. Moreover, the subscript $l$ to cycle quantities denotes the cycle representative of the $l$-th community, \emph{e.g.}, $s_l$ is the cycle entropy production.

\begin{figure}
  \hfill\includegraphics[width=.8\linewidth]{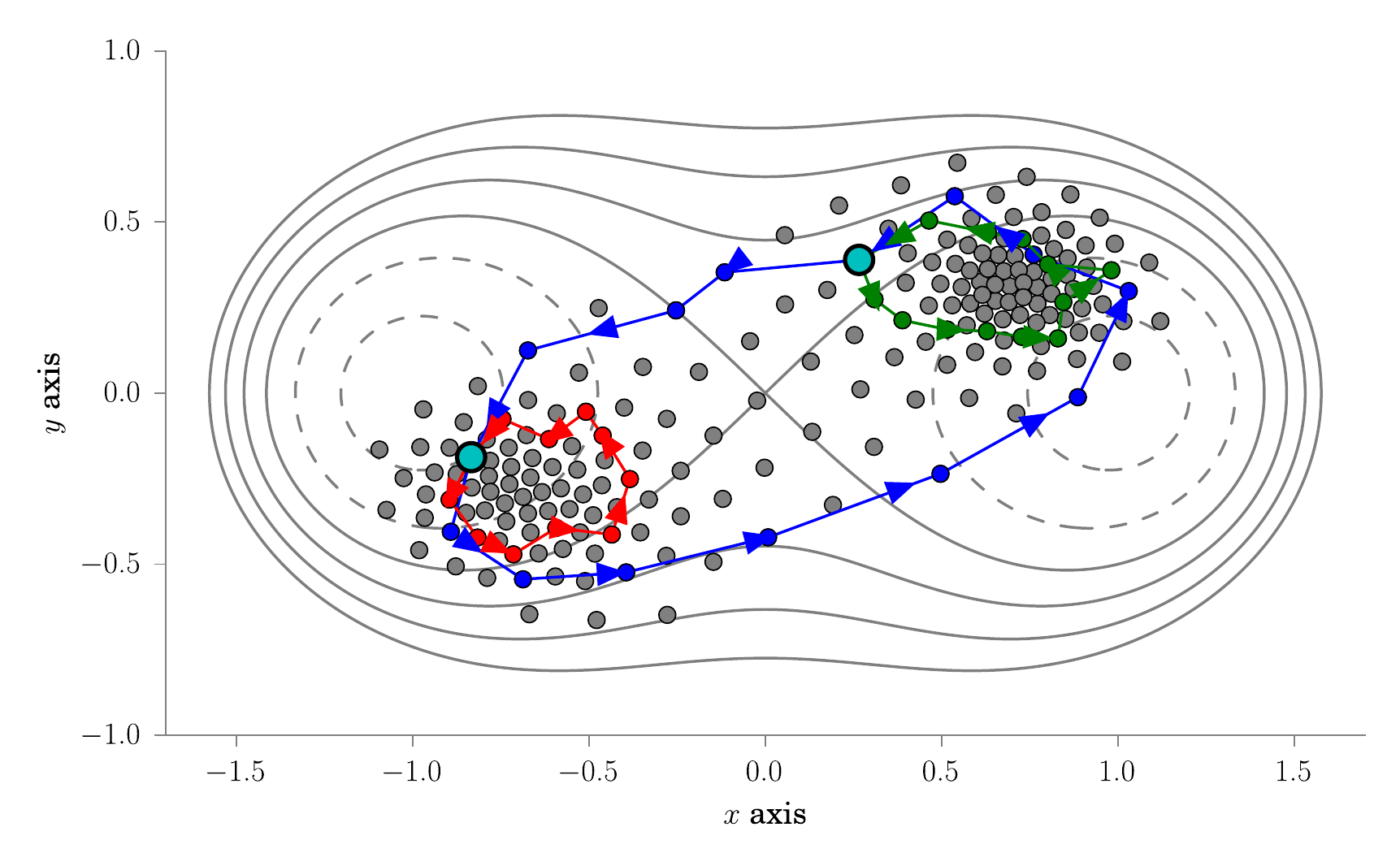}
  \caption{Illustration of the cycle representatives for each community of our model system. The centroids belonging to a representative are colored according to the communities in figure~\ref{fig:cycle_communities}, whereas the cyan states highlight the crossings of representatives. The gray points indicate the centroids not belonging to a representative, which are thus absent in the coarse-grained model.\label{fig:cycle_representative}}
\end{figure}

We now identify the \emph{new} entropy production of each representative cycle $\hat s_l$ with the entropy production of its community $S_l$. By making use of (iii), we compute the new cycle weight $\hat{\varphi}_l$ as
\begin{equation}
  \label{eq:Sl}
  S_l \stackrel{!}{=} \hat{s}_l = A_l\hat{\vhi}_l \quad\Rightarrow\quad
  \hat{\vhi}_l = \frac{S_l}{A_l} > 0.
\end{equation}
The crucial coarse-graining step consists of removing all other non-trivial cycles by setting their weights $\hat\vhi_\al$ to zero. Because the entropy production of trivial cycles is always zero, $\langle \dot{S}_{\text{tot}}\rangle$ is preserved and (i) fulfilled. All states that are not part of a community representative are thus removed with the remaining states $\hat V$ constituting the coarse-grained Markov state model, see figure~\ref{fig:cycle_representative}.

All cycles in the coarse-grained model are either a representative or trivial. We now show that the weights of the remaining trivial cycles are always positive and the coarse-grained model thus still constitutes a valid cycle-flux decomposition
\begin{equation}
  \hat\Phi^i_j = \sum_\al \hat\vhi_\al\chi^i_{j,\al}
\end{equation}
for the rescaled fluxes $\hat\Phi$, cf. eq.~\eqref{eq:decomp}. This can be achieved by demanding that all non-vanishing edge affinities $A^i_j$ are preserved. By virtue of eq.~\eqref{eq:decomp}, the new cycle weights $\hat\vhi_\al$ can then be computed from
\begin{equation}
  \exp A^i_j = 
  \frac{\Phi^i_j}{\Phi^j_i} \stackrel{!}{=}
  \frac{\hat{\Phi}^{i}_{j}}{\hat{\Phi}^{j}_{i}} = 
  \frac{\sum_{\al\in C'} \hat{\varphi}_\alpha\chi^i_{j,\alpha}}{\sum_{\al\in C'} 
    \hat{\varphi}_\alpha\chi^j_{i,\alpha}},
\end{equation}
which can be rearranged to
\begin{equation}
  \label{eq:lin_sys}
  0 = \sum_{\al\in C'} \hat\varphi_\alpha(\chi^j_{i,\alpha}
  \exp A^i_j-\chi^i_{j,\alpha}).
\end{equation}
We now pick one edge $i\ra j$ for which the edge affinity $A^i_j>0$ is positive. We then split the sum over all cycles into a sum over trivial and non-trivial cycles,
\begin{eqnarray}
  0 &= \hat\vhi_\beta(\exp A^i_j-1) + \sum_{\al\in\Cn} \hat\vhi_\al(\chi^j_{i,\al}
  \exp A^i_j-\chi^i_{j,\al}) \\
  &= \hat\vhi_\beta(\exp A^i_j-1) - \sum_l\hat\vhi_l\chi^i_{j,l}.
\end{eqnarray}
For every edge there is exactly one trivial cycle, here denoted $\beta$, for which $\chi^i_{j,\beta}=\chi^j_{i,\beta}=1$. As explained in section~\ref{sec:algo}, for positive affinity all non-trivial cycles are oriented to follow the net flow, which implies $\chi^j_{i,\al}=0$. The remaining sum over all non-trivial cycles participating in edge $i\ra j$ reduces to the weight of the representative cycles since we have set the weight of all other non-trivial cycles to zero. We have thus determined the remaining weights $\vhi_\beta>0$ of the trivial cycles, which clearly are positive.

The final step is to obtain new probabilities $\hat p_i$ for the remaining states. We simply scale out the probability of the removed states implying $p_i/p_j=\hat p_i/\hat p_j$ with normalization $\sum_{i\in V'}\hat p_i=1$. Moreover, this is sufficient to preserve the ratio of transition rates
\begin{equation}
  \exp\sigma^i_j = \frac{w^i_j}{w^j_i} = \frac{\Phi^{i}_{j}}{\Phi^{j}_{i}}
  \frac{p_j}{p_i} = \frac{\hat{\Phi}^{i}_{j}}{\hat{\Phi}^{j}_{i}}
  \frac{\hat{p}_j}{\hat{p}_i} = \frac{\hat{w}^{i}_{j}}{\hat{w}^{j}_{i}}
\end{equation}
fulfilling condition (ii). Together with the rescaled fluxes we thus obtain transition rates $\hat w^i_j=\hat\Phi^i_j/\hat p_i$, which completes the formulation of the coarse-grained Markov state model.

\subsection{Numerical test}

\begin{figure}
  \hfill\includegraphics[width=.8\linewidth]{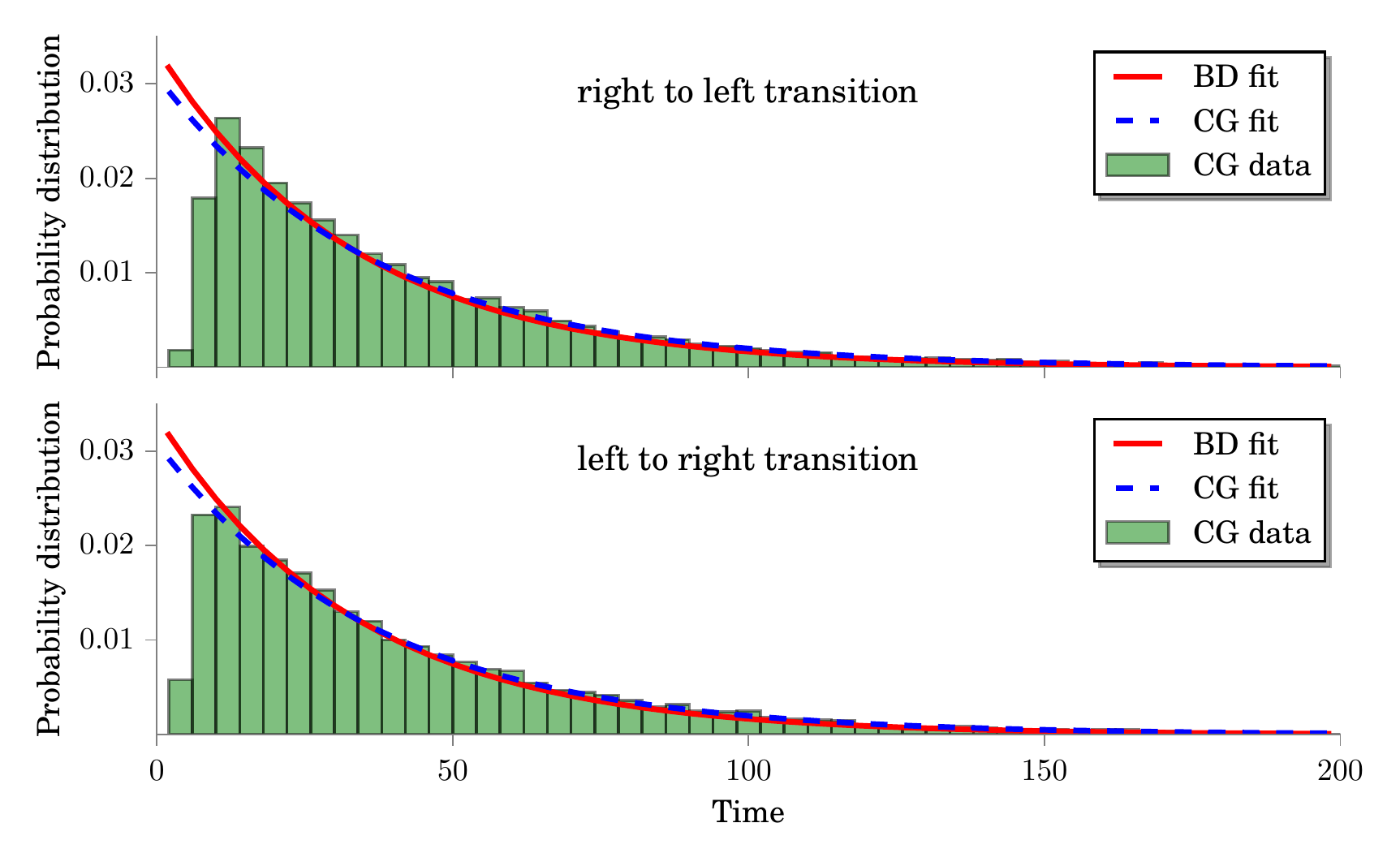}
  \caption{Rate computation for the transition between both minima. The normalized histograms show the distributions of the times needed for the specific transition, computed for the reduced Markov state model.
The blue dotted line shows the exponential fit to obtain the mesoscopic transition rate, while the red line is the fit using the histogram computed by the BD trajectory.\label{fig:histo_rates}}
\end{figure}

To test the accuracy of the coarse-grained model, we compute the forward and backward rate of traversing between both basins for the continuous BD trajectory and our reduced Markov state model. The mesoscopic rates for the reduced Markov state model have been calculated using a kinetic Monte-Carlo scheme. Figure~\ref{fig:histo_rates} shows the normalized histograms and the exponential fit (dotted blue line) obtained for the coarse-grained model as well as the exponential fit obtained from the original BD simulations (red line). Due to the symmetry of the forces, these rates should be equal, which is recovered by the coarse-grained model. Moreover, the numerical values of both rates are in excellent agreement with the BD results (see table~\ref{tab:rates}), illustrating that our coarse-graining method indeed preserves the mesoscopic time scales of the BD simulation.

\begin{table}[!h]
  \caption{Mesoscopic transition rates for the model system for the original Brownian dynamics (BD) and the coarse-grained (CG) model.}
  \begin{tabular}{l|c|c|}
    &BD&CG\\\hline
    left to right &0.029(1)&0.028(4)\\\hline
    right to left &0.029(5)&0.028(3)\\
  \end{tabular}
  \label{tab:rates}
\end{table}


\section{Critical remarks and conclusion}

There are at least two steps within the approach described here that will require further clarification and investigation. The first issue is how to find optimal cycle communities. Detecting communities on a graph is, in principle, already a challenge on its own as the vast number of heuristic approaches have shown in the past. Also, it is still a matter of discussion whether the detected communities ought to fully or only partially partition the graph. The latter is also known as fuzzy partitioning, the advantage of which is that not all cycles are assigned to a specific community as some might be matching only poorly with any of the communities. To further improve the community detection it is possible to define the link strengths $\Omega_{\alpha\beta}$ given in eq.~\eqref{eq:connectivity} somewhat differently by assigning directions to the cycle edges. In any case, once obtained the cycle communities should be checked for consistency. Of course, it is not possible to visualize the cycles by plotting them in configuration space for more than two dimensions. One way would be to project the cycles onto suitable reaction coordinates.

The second issue is how to find suitable community representatives. Here we have formulated an optimization problem with an objective function, eq.~(\ref{eq:representative}), but it might be worthwhile to also explore other strategies. It seems to be of particular importance that the cycle affinities of the representatives are, in some way, representing the averaged community affinity and their internal time scales since they will govern the mesoscopic time scales of the coarse-grained model. Especially the latter is
part of our ongoing investigation.

To conclude, we have presented a simple and scalable algorithm to construct a Markov state model in non-equilibrium that preserves entropy production and cycle affinities and, therefore, the characteristics of macroscopic non-equilibrium transport.


\ack

We acknowledge financial support by the DFG through the collaborative research center TRR 146 (project A7).


\section{Appendix}

\subsection{Cycle-flux decomposition for a simple graph}
\label{sec:cycle:decomp}

\begin{figure}
  \hfill\includegraphics[width=.8\linewidth]{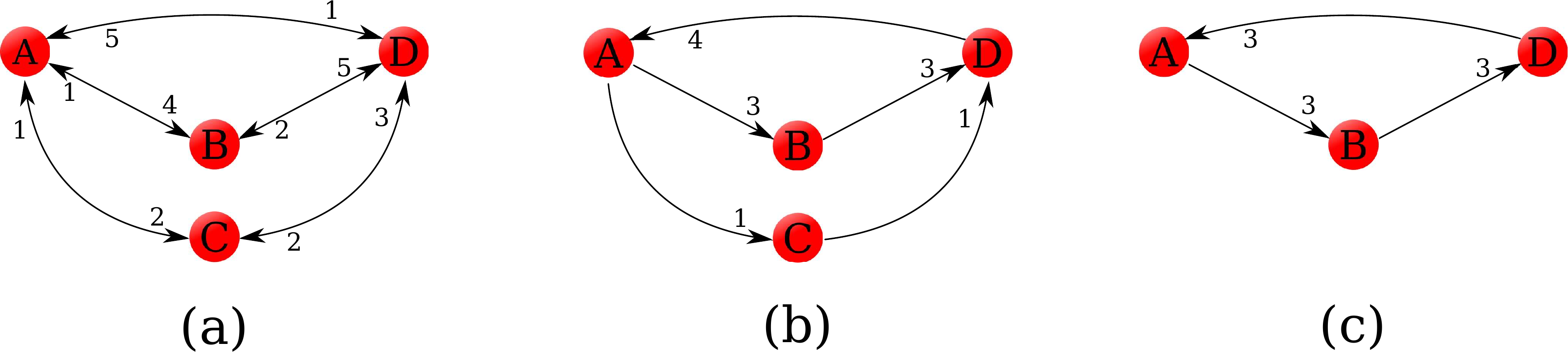}
  \caption{Example for the cycle-flux decomposition. The numbers next to the arrows are the numerical values for the fluxes $\tilde\Phi$ along edges: (a)~Initial graph with $\tilde\Phi=\Phi$, (b)~after removing the trivial cycles, and (c)~last remaining cycle.\label{app:cycle_decomp}}
\end{figure}

Here we want to give an example how to compute the cycle decomposition described in section~\ref{sec:algo}. In figure~\ref{app:cycle_decomp}(a) a simple, reversible and connected graph is shown with the vertices $A,B,C,D$. 
The arrows represent the edges and the numbers corresponding to the fluxes flowing in and out of each state. For example, the probability flux $\Phi^A_B = 4$ and the reverse flux $\Phi^B_A = 1$. It is easy to check that Kirchhoff's current law is valid, \emph{e.g.}, the flux into state $A$ equals the flux flowing out, $\Phi^{\text{in}}_A = 5+1+1 = 7 = 1+4+2=\Phi^{\text{out}}_A$. According to the Betti number, the maximal number of cycles needed for the cycle-flux decomposition is $\leq 7$. In the first step all trivial cycles (the detailed-balance part of the graph) are subtracted (see figure~\ref{app:cycle_decomp}(b)). All trivial cycles found and their weights $\varphi_\alpha$ are listed in table~\ref{tab:cycle_decomp}. The only remaining cycles are $\{A,C,D,A\}$ and $\{A,B,D,A\}$, which are subtracted in step (b) and (c), respectively. Overall, 7 cycles are needed to complete the decomposition but only the last two cycles contribute to the mean entropy production, which is, according to eq.~\eqref{eq:entropy_production}, given by
$\langle\dot{S}_{\text{tot}}\rangle = 1\cdot\ln{15} + 3\cdot\ln{50} \approx 6.62$.

\begin{table}[h!]
  \caption{Cycles and cycle weights for the graph shown in figure~\ref{app:cycle_decomp}.}
  \label{tab:cycle_decomp}
  \begin{tabular}{c|c|c|c}
    step&cycles & weights $\varphi_\alpha$& cycle affinity $A_\alpha$\\
    \hline\hline
    (a)&$\{A,B,A\}$&1&0\\
        &$\{A,C,A\}$&1&0\\
        &$\{A,D,A\}$&1&0\\
        &$\{B,D,B\}$&2&0\\
        &$\{C,D,C\}$&2&0\\
    \hline
    (b)&$\{A,C,D,A\}$&1&$\ln{15}$\\\hline
    (c)&$\{A,B,D,A\}$&3&$\ln{50}$\\
  \end{tabular}
\end{table}

\subsection{Contributing cycles}
\label{sec:cycle:det}

In this appendix, we explain how to effectively obtain the cycles that are used for the cycle-flux decomposition without first finding all possible cycles. We make use of a widely-used method in graph theory~\cite{schnakenberg1976network}. For every connected graph $G(V,E)$ a spanning tree $T(G)$ can be constructed, which is the set $V$ of all vertices connected by $|V|-1$ edges. We further demand that the edge directions in $T(G)$ are the same as in $G$. The $\nu$ edges of $G$ that are not part of $T(G)$ are called chords, with $\nu=|E|-|V|+1$. Again, it is important to preserve the original direction of each chord.

After subtracting all trivial cycles, the new flux field $\tilde\Phi$ is obtained, which contains only irreversible edges. For this graph the spanning tree $T(\tilde\Phi)$ is constructed. Adding one chord completes one non-trivial cycle. We determine all non-trivial cycles corresponding to the chords and again apply the cycle-flux decomposition as outlined in section~\ref{sec:algo}, leading to a new flux field $\tilde\Phi$. If fluxes remain, we repeat this procedure by which iteratively new cycles are added.

The complete algorithm reads as follows:
\begin{enumerate}
 \item Apply cycle-flux decomposition for all trivial cycles.
 \item Create oriented spanning tree $T(\tilde\Phi)$.
 \item Identify directed chords and find corresponding cycles.
 \item Apply cycle-flux decomposition to the newly identified cycles.
 \item Check if remaining flux field $||\tilde\Phi||_\text{max}<\text{threshold}$, otherwise repeat with step (ii).
\end{enumerate}


\section*{References}

\providecommand{\newblock}{}

\end{document}